\documentclass[10pt,twocolumn]{article}

\usepackage[activate={true,nocompatibility},final,tracking=true,kerning=true,spacing=true,factor=1100,stretch=10,shrink=10]{microtype}
\usepackage{authblk} 
\usepackage{siunitx}
\usepackage{amsmath,bm}
\usepackage{graphicx}
\usepackage{cite}
\usepackage{float}
\usepackage[all]{nowidow}
\usepackage[font=small]{caption}
\usepackage[left=48pt, right=42pt,top=46pt,bottom=60pt,headheight=15pt,headsep=10pt,letterpaper,twoside]{geometry}
\usepackage{titlesec}

\titleformat{\section}
  {\centering\normalfont\fontsize{12}{15}\bfseries}{\thesection}{.5em}{}

\title{Spatially resolving amplitude and phase of light with a reconfigurable photonic integrated circuit}

\author[ 1]{Johannes Bütow}
\author[ 1,2,3]{Jörg S. Eismann}
\author[ 4,5]{Maziyar Milanizadeh}
\author[ 4]{Francesco Morichetti}
\author[ 4]{Andrea Melloni}
\author[ 6]{David A.B. Miller}
\author[ 1,2,3,$^*$]{Peter Banzer}
\affil[1]{Institute of Physics, University of Graz, NAWI Graz, Universitätsplatz 5, 8010 Graz, Austria}
\affil[2]{Max Planck Institute for the Science of Light, Staudtstr. 2, 91058 Erlangen, Germany}
\affil[3]{Institute of Optics, Information and Photonics, University Erlangen-Nuremberg, Staudtstr. 7/B2, 91058 Erlangen, Germany}
\affil[4]{Dipartimento di Elettronica, Informazione e Bioingegneria-Politecnico di Milano, Milano 20133, Italy}
\affil[5]{Advanced Electronics and Photonics Research Centre, National Research Council Canada, Ottawa, ON, Canada}
\affil[6]{Ginzton Laboratory, Stanford University, 348 Via Pueblo Mall, Stanford, California 94305, USA}
\affil[*]{peter.banzer@uni-graz.at}
\date{} 

\begin{document}
\maketitle
\textbf{Abstract:} Photonic integrated circuits (PICs) play a pivotal role in many applications. 
Particularly powerful are circuits based on meshes of reconfigurable Mach-Zehnder interferometers as they enable active processing of light.
Various possibilities exist to get light into such circuits.
Sampling an electromagnetic field distribution with a carefully designed free-space interface is one of them.
Here, a reconfigurable PIC is used to optically sample and process free-space beams so as to implement a spatially resolving detector of amplitudes and phases.
In order to perform measurements of this kind we develop and experimentally implement a versatile method for the calibration and operation of such integrated photonics based detectors.
Our technique works in a wide parameter range, even when running the chip off the design wavelength.
Amplitude, phase and polarization sensitive measurements are of enormous importance in modern science and technology, providing a vast range of applications for such detectors.

\section{Introduction}
Photonic integrated circuits (PICs) continue to increase in overall size and complexity. Research and applications around photonic chips are becoming more widespread and advanced \cite{Chen.2018,Smit.2019}.
In particular, programmable photonic integrated circuits are increasingly deployed to actively process light on photonic chips \cite{Harris.2018,Bogaerts.2020}.
They enable the accurate control of light within reconfigurable cascaded interferometers, e.g., meshes of on-chip Mach-Zehnder interferometers (MZIs).
This approach paves the way for numerous applications, many of which are essentially impossible with conventional optics.
Applications range from arbitrary matrix operations and quantum information processing \cite{Shadbolt.2012,Miller.2013b,Carolan.2015,Harris.2016,Harris.2017,Qiang.2018,Taballione.2019,Taballione.2021} to mode sorting and communication \cite{Annoni.2017,Miller.2019}, optical beam coupling and shaping \cite{Miller.2013,Miller.2013c,Ribeiro.2016,Heck.2017}, and more.
Another possible application is the analysis of amplitudes and phases of light fields \cite{Miller.2020}.

Starting with a programmable PIC, a free-space detector can be constructed by adding an input interface, e.g., multiple gratings couplers. These grating couplers facilitate the coupling of free-space light to the integrated circuit.
The resulting spatially resolving detector consists of pixels (grating couplers), sensitive to the field polarization, and a processing unit (a programmable mesh of MZIs) capable of measuring relative amplitudes and phases. 
Combining the measurement of various field parameters in a spatially resolved fashion into one single and all-integrated platform would allow for applications reaching far beyond the limited capabilities of existing detector technology.
Examples are process surveillance in manufacturing \cite{ShihSchonLin.2004} and fiber based imaging \cite{Cizmar.2012}.
Also in nanometrology, novel methods \cite{Wozniak.2015,Picardi.2019,VazquezLozano.2019} introduced recently would greatly benefit from polarization, phase and intensity sensitive detectors enabling an unambiguous distinction of multipolar contributions to nanoparticle scattering.
Here, even with a limited number of pixels, i.e., small meshes of MZIs, a wide range of applications is conceivable.
Moreover, the scalability of programmable integrated circuits pave the way towards future detectors with many pixels \cite{Bogaerts.2020}. 
Such detectors, like most optical instruments for detecting intensity, phase and/or polarization, can only be used for quantitative measurements if they are thoroughly calibrated. In this context, Miller \cite{Miller.2020} has recently proposed a convenient theoretical scheme for the calibration of ideal meshes of MZIs. 

In this manuscript we use a reconfigurable PIC as a spatially resolving detector of amplitude and phase distributions. 
It samples free-space light beams and processes coupled fields within its mesh of MZIs.
We propose, implement and prove experimentally a novel approach with regards to both the calibration of such a detector and the analysis of unknown light beams in terms of their amplitude and phase distributions.
Our calibration technique is based on illumination with a single paraxial reference beam and, it characterizes all relevant parameters of the on-chip components during calibration, and it can deal particularly well with strong deviations from their design values.
After an introduction to the architecture of the detector in Section~\ref{sec:2} we discuss the calibration of the photonic system in Section~\ref{sec:3}. Lastly, we perform a set of proof-of-principle amplitude and phase measurements for paraxial free-space beams in Section~\ref{sec:4}.

\section{On-chip architecture} \label{sec:2}
\begin{figure*}[ht]
\centering\includegraphics[width=1\linewidth]{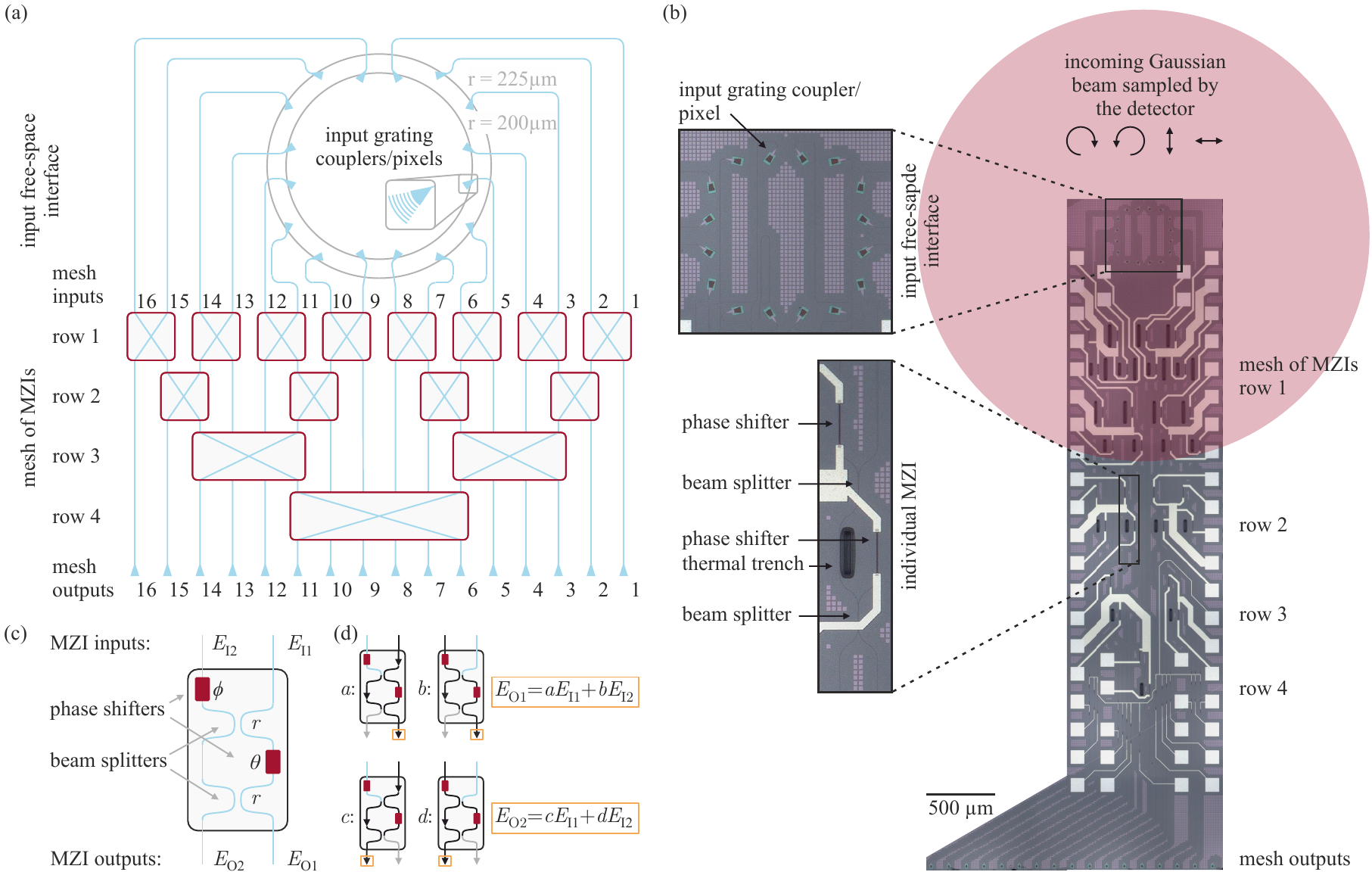}
\caption{(a)~Schematic layout of the mesh architecture. 
Grating couplers are arranged on two concentric rings in the input free-space interface.
They couple incoming free-space light into waveguides on the chip.
These are routed into a mesh of MZIs.
The mesh output consists of a set of grating couplers that send the light back into free-space after the on-chip processing.
(b)~Microscopy image of the photonic circuit with Gaussian input beam sampled by the free-space interface. Red area corresponds to FWHM of $\SI{3.3}{\milli{}m}$ of the beam on the chips interface.
(c)~Detailed structure of an individual on-chip MZI. $E_\text{I1}$, $E_\text{I2}$, $E_\text{O1}$ and $E_\text{O2}$ denote the input and output fields. $\phi$ and $\theta$ are the phase shifts induced by the phase shifters and $r$ is the field reflectivity of the beam splitters. 
(d)~Different paths the input fields can take through an MZI and their resulting contribution to the output fields, according to the mathematical description in Eq.~\eqref{eq:indivMZI}.}
\label{fig:MeshAndMZI}
\end{figure*}

The photonic chip is fabricated on a \SI{220}{\nano{}m} commercial silicon-on-insulator platform from AMF, Singapore. 
We use a custom arrangement of grating couplers, acting as pixel-like elements, to couple light into single-mode channel waveguides [Fig.~\ref{fig:MeshAndMZI}~(a)]. 
The type of grating couplers used here are usually employed to interface integrated circuits to fiber arrays \cite{Ribeiro.2016,Annoni.2017}. Nevertheless, they can be utilized for free space light fields as well \cite{Miller.2013,Ribeiro.2016,Miller.2020,Milanizadeh.2021b,Milanizadeh.2021}.
By design, they are sensitive to light polarized along their respective grating directions.
In principle, almost arbitrary numbers, distributions, and local orientations of grating couplers can be realized, resulting in a variety of different free-space interfaces.
Also the design and architecture of the individual grating couplers could be optimized or tailored for more efficient coupling with free-space beams \cite{Su.2018}.

It is worth noting here that our calibration and measurement techniques, discussed in detail below, are also applicable to an almost arbitrary pixel number and arrangement. 
However, our study focuses on a detector with $N=16$ pixels [Fig.~\ref{fig:MeshAndMZI}~(a),~(b)] placed along two concentric circles of radii \SI{200}{\micro{}m} and \SI{225}{\micro{}m} respectively, arranged radially at an alternating angle of $\pm 45^\circ$ with respect to the local radial direction.
This free-space interface has been designed originally for future applications in nano-optics \cite{Lieb.2004, Eismann.2018}. 
The free-space interface is connected to a mesh of $N-1$ on-chip MZIs. 
It is arranged in a binary-tree structure consisting of successive rows of interferometers \cite{Miller.2013}. 
Each individual MZI can be employed to perform a relative measurement between light in the two waveguides entering the respective interferometer.
The first row of MZIs processes the $N$ input fields coming from the free-space interface and sampled by the grating couplers.
Subsequent rows of MZIs process one output waveguide of each MZI of the previous row, leading to a total of $\log_2N$ rows.
The other output waveguide of each MZI, called a drop-port, is guided directly to the end of the mesh without further interaction.
The light from all waveguides reaching the end of the mesh is detected, a task that can be realized by different means.
Light can be measured directly on chip with built-in photodiodes or transparent photoconductors \cite{Morichetti.2014}.
Alternatively another set of grating couplers can terminate the mesh output waveguides, coupling the light out of the photonic chip again and allowing detection in free-space, e.g., with a camera or by using fiber bundles and photodiodes. As shown in Fig.~\ref{fig:MeshAndMZI}~(a),~(b), we use free-space detection for this demonstration.

The spatially resolved measurement (of amplitude and phase) of an input light field sampled with the free-space interface becomes possible by a pair-wise interferometry within the individual MZIs.
Similar to its conventional equivalent, an on-chip MZI [Fig.~\ref{fig:MeshAndMZI}~(b),~(c)] consists of two input waveguides, with complex valued fields $E_\text{I1}$, $E_\text{I2}$, and two output waveguides, with their respective fields $E_\text{O1}$, $E_\text{O2}$.
In between, there are two beam splitters and two phase shifters.
The beam splitters are implemented by means of directional couplers, but other solutions like multimode interference couplers could be used as well.
The field reflectivity $r$ of such a beam splitter is then influenced by the length of the coupling section and the gap distance between the coupled waveguides. 
It is reasonable to assume $r$ to be essentially the same for all beam splitters across all MZIs in the mesh due to the high uniformity of the manufacturing processes.
The necessary phase shifters are implemented by integrating small titanium nitride heaters on top of the silicon core of the waveguide, actuated by applying a voltage. This induces a local change of the refractive index and, hence, the optical path length of a nearby waveguide, resulting in phase shifts $\phi$ and $\theta$ [see Fig.~\ref{fig:MeshAndMZI}~(c)]. More details on the technology platform can be fount in Ref.~\cite{Milanizadeh.2021}, where a different circuit topology was fabricated by using the same MZI building blocks and the same silicon photonic foundry.

All elements on the photonic chip are standard foundry elements for photonic architectures, designed for a wavelength of \SI{1550}{\nano{}m}.
Due to the position-dependent orientation of the grating couplers, it is not possible for an incoming beam to impinge onto all couplers at their design angle of incidence ($12^\circ$ relative to normal incidence on the chip).
Solutions to implement perfect vertical surface grating couplers in silicon waveguides have been recently proposed using subwavelength design engineering \cite{KamandarDezfouli.2020}.
In the present work, a solution affecting all couplers in the same way, is to send the light onto the chip at normal incidence, redshifting the optimum wavelength of operation.
We thus operate the system at \SI{1600}{\nano{}m}.
This also affects other components of the system, specifically the reflectivity of the on-chip beam splitters, which are designed for 50:50 splitting at \SI{1550}{\nano{}m}.
This is, however, not a problem since following our method of calibration and operation of the detector, neither one requires specific splitting ratios from the beam splitters.
All properties of the on-chip components are, in fact, accounted for and determined during the calibration of the chip.

\section{Calibration of the photonic mesh and on-chip components} \label{sec:3}
A careful calibration of all on-chip elements, specifically the mesh of MZIs, is crucial before the system can be utilized to measure field distributions impinging on the free-space interface of the photonic chip. 
The calibration is based on a known incoming field.
The resulting waveguide input fields act as phase \cite{Miller.2020} and amplitude reference.
The influence of the mesh on the reference input is recorded by measuring the intensities of the mesh outputs while, at the same time, sweeping phase shifters of MZIs through a range of values.
The parameters of the system that need calibration can afterwards be obtained by fitting the resulting intensities to a transmission function containing the unknown properties of the mesh as free parameters.
The reflectivity of the beam splitters and possible imbalances between the input amplitudes are important parameters to be characterized. 
The latter depend on the input field distribution and can further be affected by imperfections in the grating couplers or the waveguides arising during fabrication.

\begin{figure}[ht]
\centering\includegraphics[width=1\linewidth]{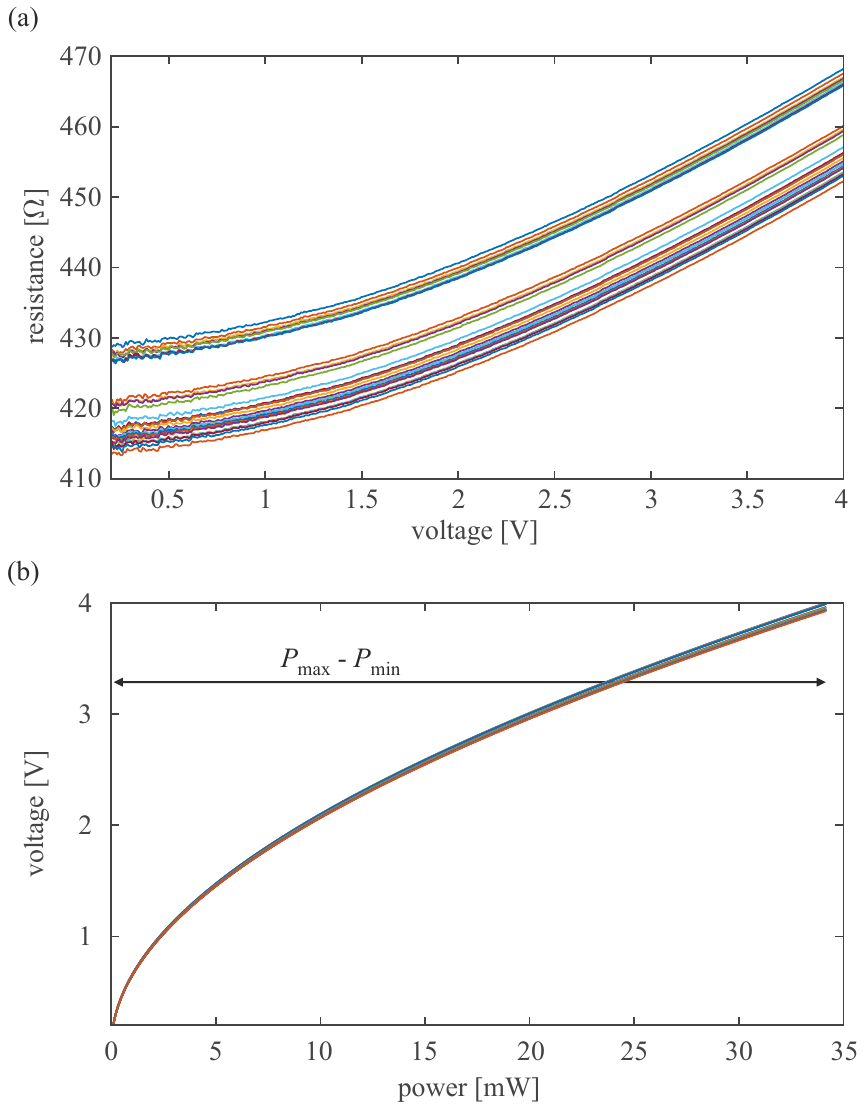}
\caption{(a)~Electrical resistance of every phase shifter within the mesh of MZIs as function of the applied voltage. (b)~Inverted characteristics $V(P)$ to identify phase shifter specific drive voltages resulting in a common and linear increase in power.}
\label{fig:phaseshifters_resistance}
\end{figure}

The most crucial unknowns in the system that require calibration are the relations between the actual phase shifts $\phi_k$ and $\theta_k$, and the applied voltages $V$ to every individual phase shifter ($k = 1,...,N-1$).
In good approximation, the phase shift of the integrated heaters is linearly dependent on the applied power $P=V^2/R$.
The resistance $R$ of every phase shifter is also a function of the applied voltage, presumably because the resistance changes with temperature. 
We therefore measure the $R(V)$ characteristics of every phase shifter individually for voltages between 0.2 and 4\,V, see Fig.~\ref{fig:phaseshifters_resistance}~(a).
The corresponding $P(V)$ characteristics are used to, firstly, identify a common minimum and maximum power ($P_\text{min}$ and $P_\text{max}$), accessible within the given voltage range.
Secondly, their inverted characteristics $V(P)$ can be used to identify phase shifter specific drive voltages that, when applied, result in a common and linear increase of power between $P_\text{min}$ and $P_\text{max}$ [Fig.~\ref{fig:phaseshifters_resistance}~(b)].
The resulting phase shifts $\phi_k$ and $\theta_k$ are then increasing linearly with the applied power and cover a phase range $\gamma_\text{range}$, which is the same for all phase shifters and is left as parameter to be determined.
Additionally, path-length differences between waveguides can have a significant effect on the relative phases within the mesh.
Waveguide sections have been designed in order to share the same geometrical length but due to fabrication imperfections they can add random (yet fixed) relative phase shifts within the mesh. 
Also, an imperfect alignment of the calibration beam can introduce such additional phase values to the fields entering the first row of MZIs.
These overall added phase shifts are taken into account during calibration by attributing them as phase offsets $\phi_k^\text{off}$ and $\theta_k^\text{off}$ to the individual phase shifters.
Ultimately, the phase shifts can be written as
\begin{equation}
    \phi_k = \frac{P-P_\text{min}}{P_\text{max}-P_\text{min}} \cdot \gamma_\text{range} + \phi_k^\text{off} 
\end{equation}
and
\begin{equation}
    \theta_k = \frac{P-P_\text{min}}{P_\text{max}-P_\text{min}} \cdot \gamma_\text{range} + \theta_k^\text{off}
\end{equation}
which leaves leaves a total of $[1+ 2\cdot k]$ unknown parameters describing the phase shifts applied within the mesh.

Analyzing the effect of the mesh on a reference input requires modeling the mesh output intensities for an arbitrary choice of mesh parameters and input fields.
The output fields $E_\text{O1}$, $E_\text{O2}$ of an individual MZI can be calculated in a convenient manner by using 2x2 unitary matrices acting on two input fields $E_\text{I1}$, $E_\text{I2}$:
\begin{equation} \label{eq:indivMZI}
    \begin{pmatrix}
    E_\text{O1} \\
    E_\text{O2} 
    \end{pmatrix}
    = 
    \begin{pmatrix}
    a & b\\
    c & d
    \end{pmatrix}
    \begin{pmatrix}
    E_\text{I1} \\
    E_\text{I2} 
    \end{pmatrix},
\end{equation}
with\begin{align*}
    a &= r^2  \exp{(\text{i}\theta)} + t^2,\\
    b &= (tr  \exp{(\text{i}\theta)} + rt) \exp{(\text{i}\phi)},\\
    c &=  rt\exp{(\text{i}\theta)} + tr,\\
    d &= (t^2\exp{(\text{i}\theta) + r^2)} \exp{(\text{i}\phi)},
\end{align*}
where $t = \text{i}\sqrt{1-r^2}$. 
A detailed derivation and explanation of this formalism can be found in \cite{Miller.2020}.
Fig.~\ref{fig:MeshAndMZI}~(d) shows the different paths possible for the input fields and how they relate to the entries $a,b,c$ and $d$.

\begin{figure}[htbp]
\centering\includegraphics[width=1\linewidth]{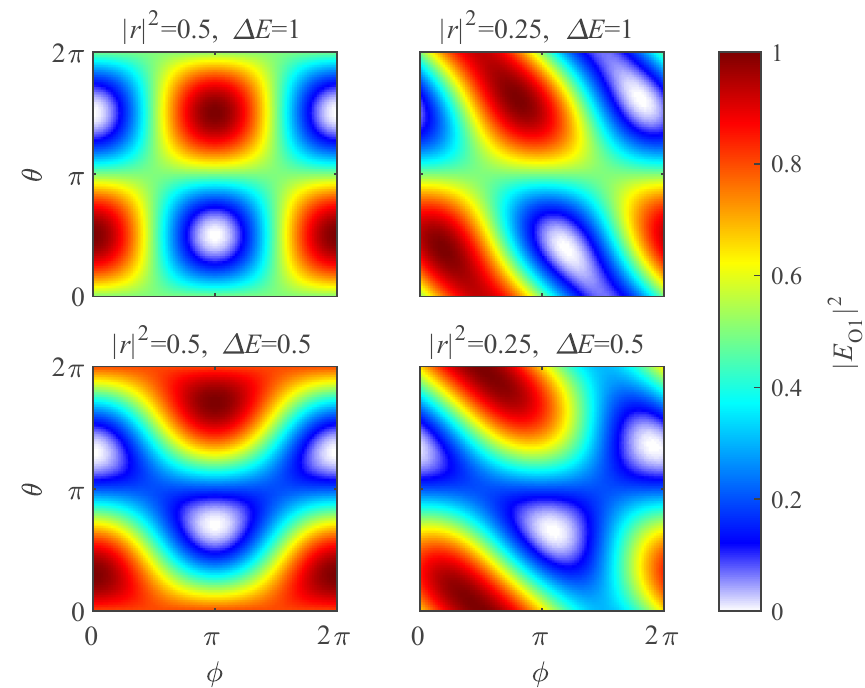}
\caption{Normalized, single MZI DIMs for in-phase inputs prior to the interferometer. Illustrated are DIMs for different beam splitter power reflectivities $|r|^2$ and input amplitude ratios $\Delta E$.}
\label{fig:InsightsMZI}
\end{figure}

In Fig.~\ref{fig:InsightsMZI}, we visualize the output values $|E_\text{O1}|^2$ of an exemplary theoretical MZI, while varying the phase shifts $\phi$ and $\theta$, here from $0$ to $2\pi$. This is referred to as drop-port intensity map (DIM) of an MZI or a mesh output respectively.
Fig.~\ref{fig:InsightsMZI} shows several of such (normalized) DIMs for different values of the beam splitter power reflectivity $|r|^2$ and input fields with an amplitude ratio $\Delta E = |E_\text{I1}|/|E_\text{I2}|$ and a fixed in-phase relation before entering the MZI. 
Four cases for different reflectivities (left: $|r|^2 = 0.5$; right: $|r|^2 = 0.25$) and amplitude ratios $\Delta E$ (top: $\Delta E = 1$; bottom: $\Delta E = 0.5$) are shown.
The maps change continuously if the parameters in Eq.~\eqref{eq:indivMZI} are changed continuously.
The DIM can thus be correlated with these parameters, and a least-squares fitting routine can be used to retrieve $r$ and $\Delta E$ (as well as other parameters) from the corresponding DIMs.
It is worth noting how a phase difference $\Delta \phi$ between the input fields (before entering the MZI) would translate the DIMs horizontally by an amount $\Delta \phi$. This conveys intuitively how, with a calibrated MZI, the relative phase between input fields can be inferred from this horizontal translation of the DIM.

Generally, these MZI meshes can be analyzed, and the overall transmission matrix that represents them constructed, using some simplifying topological results \cite{Pai.2020}. Any such mesh in which light flows only ``forwards'' (e.g., light flowing from inputs to outputs as in our mesh in Fig.~\ref{fig:MeshAndMZI}~(a) corresponds topologically to a directed acyclic graph. As shown in \cite{Pai.2020}, such a graph can be factored into a set of successive topological ``columns'' of MZIs in which the fields in the different MZIs do not interact within the “column”. These topological “columns” correspond to the ``rows'' in our description. The transmission matrix representing one such ``column'' (or our ``row'') can then be written as a block diagonal matrix with the corresponding 2x2 transmission matrices positioned appropriately on the diagonal (and with zeros in all other off-diagonal entries). Different matrix columns can then correspond to the different input waveguides of the MZIs in such a topological ``column'', and different matrix rows can correspond to the different output waveguides of these same MZIs. The transmission matrix for the first MZI row (topological ``column'') of our mesh can therefore be written as
\begin{equation} \label{eq:}
\stackrel{\leftrightarrow}{\mathbf{M}}_1 = 
\begin{pmatrix}
  a_1 & b_1 & 0 & \dots &  0  & 0 \\
  c_1 & d_1 & 0 & \dots &   0 & 0 \\
  0   & 0   & \ddots &  &\vdots & \vdots \\
  \vdots&\vdots & &  \ddots &0 &0\\
  0 & 0 & \dots  &0 & a_8&b_8\\
  0 & 0 & \dots  &0 &c_8&d_8
  \end{pmatrix} \ .
\end{equation}
with the subscripts 1, 2, etc. corresponding to the different MZIs in the mesh row (or topological ``column''). The $N \times N$ matrices for subsequent rows of our mesh are constructed similarly. If there is no MZI on a given waveguide or waveguide pair in a given mesh row (topological ``column''), the corresponding diagonal elements of the matrix for this mesh row are set to 1. The matrix for the entire mesh is then simply the product, in order, of all of these $N \times N$ matrices \cite{Pai.2020} for the different mesh rows.

To obtain a set of waveguide input fields for the calibration of the photonic device, a paraxial reference beam acting as phase and amplitude reference is sent at normal incidence onto the free-space interface of the photonic architecture [Fig.~\ref{fig:MeshAndMZI}~(b)].
The size of the beam is chosen sufficiently large (FWHM of $\SI{3.3}{\milli{}m}$ on the chip surface, much grater than the diameter of the rings of grating couplers that form the free-space interface) to provide a quasi-planar phase front. Further, the beam is (right-handed) circularly polarized to ensure efficient coupling to every grating coupler within the free-space interface independent of their respective orientations.
It is worth noting that that due to these different orientations, the circularly polarized calibration beam will not result in 16 in-phase input fields in the waveguides.
However, this geometrical phase factor can be calculated in a straight-forward manner and is accounted for in the calibration.

\begin{figure*}[htbp]
\centering\includegraphics[width=1\linewidth]{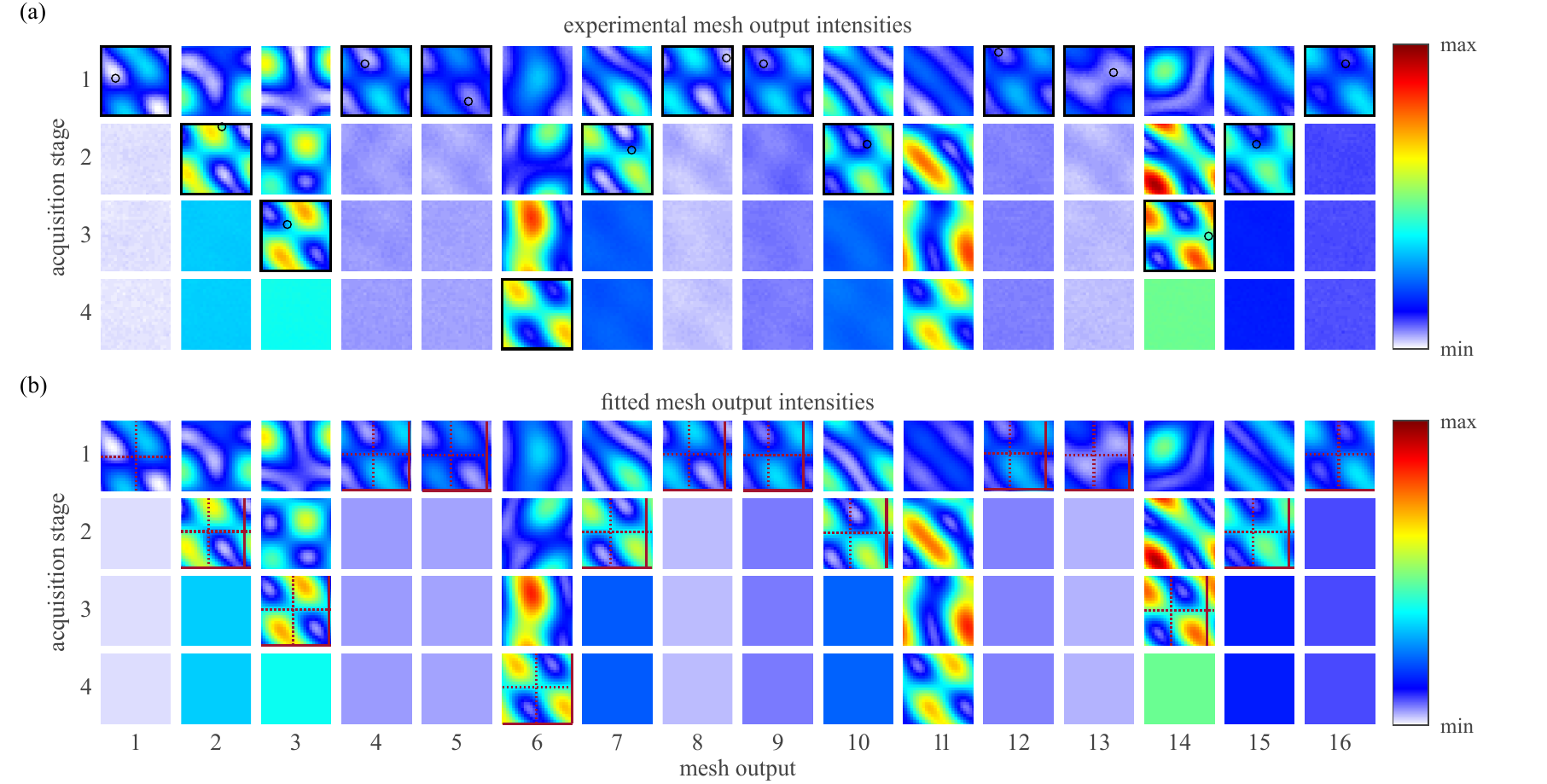}
\caption{Calibrating the photonic chip. All 16 mesh outputs are shown for each acquisition stage.  
        (a)~Experimentally recorded values. Maps with black frames correspond to MZIs of the particular row in the mesh that were sampled in the corresponding acquisition stage. The MZI voltage setting that was chosen as a fixed value for subsequent acquisition stages is marked by black circles.
        (b)~Theoretical fit, emulating the experiment.
        MZIs belonging to rows corresponding to the respective acquisition stage are superposed with solid/dotted red lines corresponding to phase shifts of even/odd multiples of $\pi$ respectively, according to the results of the calibration.
        }
\label{fig:cali_experiment_and_theory}
\end{figure*}

The recorded data is based on a simultaneous readout of all mesh output intensities.
Progressively, each row of MZIs is actively controlled to sample a grid of phase shifter drive voltages whilst keeping all other MZIs of the mesh at a constant setting.
This results in four acquisition stages for the calibration, each showing 16 intensity maps [see Fig.~\ref{fig:cali_experiment_and_theory}~(a)].
Experimentally, this data is recorded by imaging the output grating couplers of the mesh on a camera. Regions around the individual grating couplers are integrated while applying voltages to the phase shifters. The temperature of the photonic chip is actively stabilized to $26^\circ$\,C.
During the first acquisition stage, all MZIs belonging to row 1 are sampling their respective grid of $25 \times 25$ voltages in parallel. 
The voltages for MZIs of row 2-4 are kept at constant values, in this case at the lowest values of their corresponding voltage grid (around 0.2\,V). 
The 16 mesh outputs are recorded simultaneously, resulting in a grid of $25\times 25$ intensities for each output. They are shown in the first row of Fig.~\ref{fig:cali_experiment_and_theory}~(a).
DIMs corresponding to MZIs actively swept during the acquisition stage are marked with black frames.
Before the second acquisition stage can begin, the MZIs of the first row are set to specific constant voltage values.
These are chosen based on the measured intensities during stage 1 to send similar intensities into the subsequent row 2 of MZIs.
These voltage settings corresponding to sets of phases are marked with black circles in Fig.~\ref{fig:cali_experiment_and_theory}~(a), and they fix the intensities measured at these DIMs during further acquisition to a constant value.
Deviations from these constant intensities might arise either due to limitations in the imaging of the output grating couplers (introducing a small cross-talk between the recorded intensities) or other reasons such as thermal cross-talk.
The measuring procedure is then repeated for acquisition stages 2-4.

We can now perform a least-squares fitting routine by using the experimentally recorded output intensities, a reference in-phase input vector and a transmission matrix describing the mesh of MZIs, resulting in corresponding theoretical maps.
For this purpose the MATLAB function ``lsqcurvefit'' with the algorithm ``trust-region-reflective'' was used within the scope of this manuscript.
The fit uses the mesh and input parameters requiring calibration as free parameters for the optimization.
These are the 30 individual offsets of the phase shifters in the mesh ($\phi_k^\text{off}$ and $\theta_k^\text{off}$), the common phase range $\gamma_\text{range}$ for the phase shifters, an overall beam splitter field reflectivity $r$ and 16 individual mesh input amplitudes. 
We also allow for offsets in the recorded intensities. 
Intensity offsets might arise due to incoherent back reflections in the circuit or scattered light from the photonic chips surface or edges.
They do not affect the interferometric measurements of relative amplitude and phase for which the information is not in the absolute power levels of the DIMs, but in the relative intensity profiles.
The calculated mesh output is shown in Fig.~\ref{fig:cali_experiment_and_theory}~(b).
Phase shifts corresponding to even/odd multiples of $\pi$ are indicated by red solid/dotted lines, respectively. The retrieved common phase range is identified to be $\gamma_\text{range} = 1.94\ \pi$ and the overall value of the reflectivity is $|r|^2 = 0.27$.

\section{Amplitude and phase measurements on paraxial beams} \label{sec:4}

\begin{figure*}[htbp]
\centering\includegraphics[width=1\linewidth]{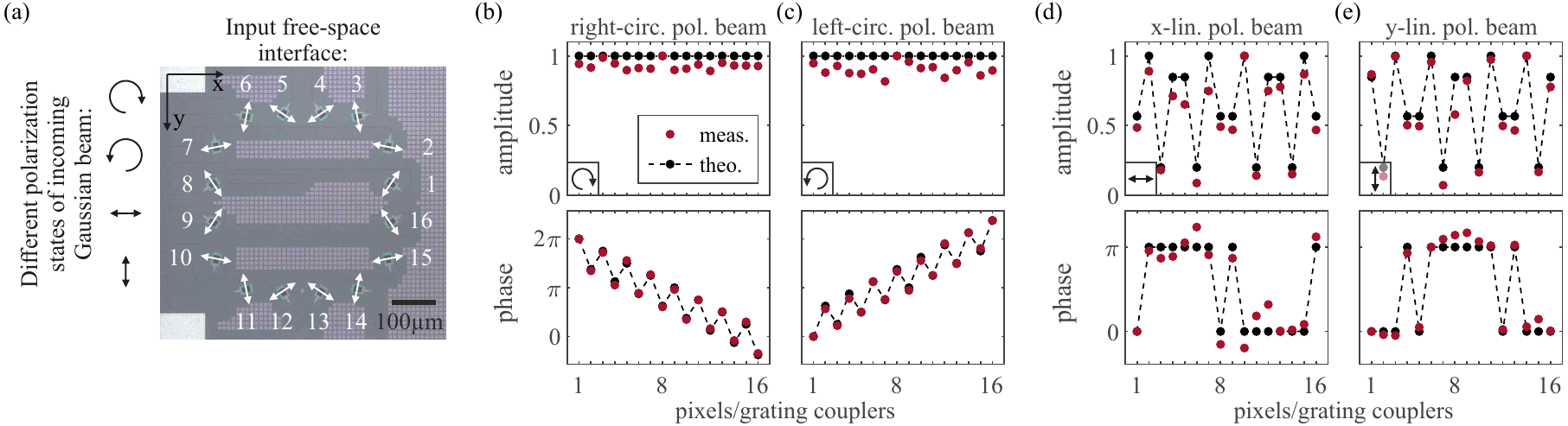}
\caption{(a)~Microscopy image of the input free-space interface, together with the different polarization states of an incoming Gaussian beam (black arrows) and the local orientations of the polarization for which the pixels/grating couplers were designed (white arrows).
(b)-(e)~Measured and theoretically expected values in red and black, respectively, for several different incoming beams sampled by the free-space interface. The investigated beams are (b) a right-handed circularly polarized beam (similar to calibration), (c) a left-handed circularly polarized beam, (d) a linearly x-polarized beam and (e) a linearly y-polarized beam. Normalized amplitudes are shown in the upper row, whereas the corresponding phases are shown in the lower row.}
\label{fig:resultsExperiment}
\end{figure*}

After successful calibration of the photonic mesh, we can now use the chip to measure unknown intensity and phase distributions impinging on the free-space interface and retrieve their relative amplitudes and phases at the positions of the grating couplers and within the polarization basis defined by their orientations [Fig.~\ref{fig:resultsExperiment}~(a)].
As a proof of principle, we measure the amplitudes and phases of four paraxial fundamental Gaussian beams with different polarization states and compare the experimental results to our theoretical expectation. 
The data acquisition for such a measurement is similar to before, progressively sampling rows of MZIs through a predefined grid of voltages whilst recording the mesh output intensities.
However, circuit parameters like the transmission matrix of the mesh (i.e., phase shift/voltage relation and MZI reflectivity) were determined during calibration.
This time, the amplitudes and phases of the input waveguide fields, resulting from the incoming beam, are of interest.
Consequently, the measurement fitting algorithm now uses a complex input vector, describing the unknown input field sampled by the couplers, as free parameter.
The result corresponds to the local amplitude and phase information of the input distribution impinging on the free-space interface and coupling into the photonic chip.

We analyze a right-handed circularly polarized fundamental Gaussian beam [Fig.~\ref{fig:MeshAndMZI}~(b)].
The number of driving voltages for the phase shifters was reduced to 10, resulting in a sampled grid of $10\times 10$ voltages for each MZI.
The retrieved amplitude and phase distribution by the free-space interface is plotted in Fig.~\ref{fig:resultsExperiment}~(b).
The local orientations of each grating coupler, i.e., pixel, introduce a geometrical amplitude and phase pattern, which is also reflected by the theoretical values. These values (connected by a dashed line for better visualization) are calculated in a straight forward manner by locally decomposing the electric field of the incoming beam into the polarization basis defined by the grating coupler orientations. 
The measured amplitudes and phases fit the theoretically predicted values very well.

Further measurements were performed for different polarization states of the incoming beam.
The results of these measurements are shown in Figs.~\ref{fig:resultsExperiment}~(c), (d) and (e), for a left-handed circularly polarized, linearly $x$-polarized and linearly $y$-polarized incoming beam, respectively.
Measurements and theory are in very good agreement for all examples shown.

\section{Summary and outlook} \label{sec:5}
In this manuscript, we have used a reconfigurable PIC as a spatially resolving detector of free-space amplitude and phase distributions.
To be able to perform measurements of this kind we have proposed, experimentally implemented and verified a method for the calibration of the system using a single reference beam.
By utilizing the calibrated photonic detector, we also presented proof-of-principle measurements of amplitude and phase distributions of different input beams within the polarization basis set by the design of the input free-space interface. 
Calibration and operation rely on the same principle of performing a multi-parameter fit between recorded mesh output intensities and a transmission matrix, which describes the complete mesh of MZIs.
The difference between them lies in the choice of free parameters to the fit.
In case of calibration, a known reference input beam is sent onto the free-space interface to calibrate the chip. Thus, free parameters stem from on-chip elements of the photonic circuit.
In contrary, for a measurement, an unknown input beam is measured by the prior calibrated and hence known system. There, the free parameters are amplitudes and phases of the incoming beam.

Compared to the approach in \cite{Miller.2020} of measuring beams, which is based on ``perfect'' MZIs with 50:50 beamsplitters (or at least ones that could be effectively perfected with the addition of significant complexity in the circuit \cite{Miller.2015}) and requires two beams and/or additional detectors for calibration, our approach, though more measurements of powers are necessary, has several advantages. 
Calibration of the photonic chip with a single paraxial beam is especially convenient experimentally and is applicable to an almost arbitrary input free-space interface design.
The calibration routine runs at the click of a button and no mechanically moving parts or other means of manipulating the reference input are required. 
Furthermore, because our approach can operate with ``imperfect'' beam splitter ratios, even quite far from 50:50, our method of calibration and operation of the system is not restricted to the design wavelength of the on-chip components.

Notably, the programmable photonic circuit with its characterized mesh of MZIs is also ready to be used for other applications involving calibrated meshes of MZIs, like, e.g., the generation of multimode optical fields.
In the future the pixel number and corresponding mesh size can be expected to grow further. 
The development of novel building blocks, especially polarization splitting grating couplers \cite{Nambiar.2018}, could further improve the capabilities of the photonic detector. Without any additional changes to the mesh of MZIs or to our method of calibration and operation, these novel pixels could be implemented in the input free-space interface, enabling spatially resolved polarization measurements.
Integration of the overall system could be improved further by the use of integrated on-chip measurements of intensity \cite{Morichetti.2014}, which would eliminate the need for additional equipment.
Lastly, other material platforms could extend the range of application of this photonic detector further, to the mid-IR range \cite{Lin.2017} even to visible light \cite{Munoz.2019}.

\vspace{.4cm}
\textbf{Funding: }
This work was supported by the European Commission through the H2020 project SuperPixels (grant 829116).
\bibliographystyle{ieeetr}
\footnotesize 
\bibliography{main}
\end{document}